# Time series modeling and large scale global solar radiation forecasting from geostationary satellites data

Cyril Voyant[1,2*], Pierrick Haurant[3], Marc Muselli[1], Christophe Paoli[4], Marie-Laure Nivet[1]

[1] University of Corsica, CNRS UMR SPE 6134, 20250 Corte, France

[2] Castelluccio Hospital, Radiotherapy Unit, BP 85, 20177 Ajaccio, France

[3] Centre Thermique de Lyon (CETHIL), UMR CNRS 5008 / INSA Lyon, France

[4] Galatasaray University, Computer Science Department, Çırağan Cad. No:36, Ortaköy 34357, Istanbul, Turkey

*corresponding author; tel +33495293666, fax +33495293797, cyril.voyant@ch-castelluccio.fr

Abstract.

When a territory is poorly instrumented, geostationary satellites data can be useful to predict global solar radiation. In this paper, we use geostationary satellites data to generate 2-D time series of solar radiation for the next hour. The results presented in this paper relate to a particular territory, the Corsica Island, but as data used are available for the entire surface of the globe, our method can be easily exploited to another place. Indeed 2-D hourly time series are extracted from the HelioClim-3 surface solar irradiation database treated by the Heliosat-2 model. Each point of the map have been used as training data and inputs of artificial neural networks (ANN) and as inputs for two persistence models (scaled or not). Comparisons between these models and clear sky estimations were proceeded to evaluate the performances. We found a normalized root mean square error (nRMSE) close to 16.5% for the two best predictors (scaled persistence and ANN) equivalent to 35-45% related to ground measurements. Finally in order to validate our 2-D predictions maps, we introduce a new error metric called the gamma index which is a criterion for comparing data from two matrixes in medical physics. As first results, we found that in winter and spring, scaled persistence gives the best results (gamma index test passing rate is respectively 67.7% and 86%), in autumn simple persistence is the best predictor (95.3%) and ANN is the best in summer (99.8%).

| | | | |
|---|---|---|---|
| $X_t(x_i, y_j)$ | X parameter concerning the pixel $(x_i, y_j)$ and the time $t$ | $f, g,$ $\omega_{ij}^1, b_i^1, \omega_i^2 + b^2, H$ and $In$ | MLP parameters concerning activation functions, weights and bias, number of hidden and input nodes |
| $n_t(x_i, y_j)$ | Clearness of the atmosphere (*unitless*) | $X_t / x = \{x_1 \ldots x_n\}$ $Y_t / y = \{y_1 \ldots y_n\}$ | Time series $X_t$ and $Y_t$, and possible associated values |
| $\rho_t(x_i, y_j)$ | Measured albedo (*unitless*) | $L^\tau$ | Lag operator and associated order |
| $\rho_t^{cloud}(x_i, y_j)$ | Albedo of the brightest clouds (*unitless*) | $H(X_t), H(X_t|Y_t)$ and $MI(X_t, Y_t)$, | Marginal / conditional entropies and the mutual information (*bit*) |
| $\rho_t^{CS}(x_i, y_j)$ | Albedo of the ground under clear sky (*unitless*) | $p(x), p(y)$ and $P(x,y)$ | Marginal and joint probabilities distribution function of $X_t$ and $Y_t$ |
| $CSI_t(x_i, y_j)$ | Clear sky index (*unitless*) | $nRMSE$ | Normalized root mean square error (*%*) |
| $I_t(x_i, y_j)$ | Global radiation (*Wh/m²*) | $r_p, r_m$ | Pixels distance (polar coordinate) concerning predicted and measured map $(=\sqrt{x_i^2 + y_i^2}, m)$ |
| $I_t^{CS}(x_i, y_j)$ | Global radiation under clear sky (*Wh/m²*) | $r(r_p, r_m)$ | Distance between pixels from predicted ($r_p$) and measured map ($r_m$) (*m*) |
| $p$ | Number of parameters used for create model | $\gamma$ and $\Gamma(r_p, r_m)$ | Gamma index and gamma score (*unitless*) |
| $\epsilon_{t+1}(x_i, y_j)$ | Prediction error (measurement-prediction) | $Tol_r$ and $Tol_I$ | Distance (also called DTA) and intensity Tolerances (*m, Wh/m²*) |
| $f_n$ | Linear or non-linear model | | |

# 1. Introduction

The production and use of non-renewable resources based on fossil fuels combustion are responsible of real public health problem and raise environmental concerns. There are lots of alternatives such as photovoltaic and wind energy sources, which one of the main advantages are the renewable and inexhaustible aspects and the main disadvantage is related to their intermittencies (Hocaoglu, 2011; Voyant et al., 2012). These non-continuities can cause a demand/production unbalance involving irrelevant wind or solar systems uses. To overcome this problem, it is necessary to predict the resource and to manage the transition between different energies sources (Bouhouras et al., 2010; Darras et al., 2012; Muselli et al., 1998b). Considering the grid manager's point of view (Köpken et al., 2004), needs in terms of prediction can be distinguished according to the considered horizon: following days, next day by hourly step, next hour and next few minutes. We choose in this paper to focus only on h+1 horizon prediction of global radiation as a first step (one hour in advance). Of course, we are aware concerning the importance of other horizons (Voyant et al., 2013b). Note that it is appreciable to know the eventual fluctuations at least 30 minutes ahead (ignition delay of turbines) for an ideal electrical grid management (Troccoli, 2010). With efficient prediction tools dedicated to grid managers, the PV part in the mix energy would be increased; actually in France, the intermittent energy contribution is limited to 30%.

Several prediction methods have been developed by experts and can be divided in three main groups: methods using mathematical formalism of times series (TS) (De Gooijer and Hyndman, 2006; Elminir et al., 2007; Hamilton, 1994), numerical weather predictions (NWP) and models based on clouds detection (Inness and Dorling, 2012; Perez et al., 2013). In this study, we have chosen to study prediction methods of the first group and we will study if this methodology can be an alternative to the NWP models. Not that in the literature, the NWP models are compared against ground measurements and the error established is approximately 30-40% but depends on the orography and micro-climate studied (Paulescu, 2013). The time series formalism (TS) and modelling is often used in 1 dimensional (1-D) global radiation predictions, i.e. related to one measurement system at ground level (Pons and Ninyerola, 2008). Persistence, autoregressive models, multilayer perceptron (MLP) and more widely artificial neural network (ANN) often applied to this aim (Hocaoglu, 2011; Mellit et al., 2009; Voyant et al., 2012).

In this paper, we will complete the first prediction results exposed in (Haurant et al., 2013) and we will show that the TS formalism applied to surface solar irradiances (SSI) estimations reduced to one point (1-D approach) can be generalized in the 2-D case, even if no ground detector is present. Alternative approaches are available in (Loyola R., 2006; Rahimikhoob et al., 2013). From this point of view, satellite derived SSI maps extracted from the HelioClim-3 (HC-3) (Rigollier et al., 2004) database and centered on Corsica are used as hourly 2-D data generator. Each data series is processed with stochastic estimators in order to generate 1158 predictions per hour (1158 pixels per map separately treated). In fact, for an overall year of prediction, it is necessary to generate more than 10 million of hourly predictions (24x365x1158). The purpose of this paper is to generate one hour in advance predicted global radiation maps of a specific area from HC-3 SSI maps. But as data used are available for the entire surface, the method can be easily generalized. The geographical effects are taken into account using clear sky index in addition to temporal or seasonal phenomena (Allan, 2011). The uncertainty of the used satellite derived SSI maps is about 16-23% (http://www.soda-

is.com/eng/helioclim/helioclim3_uncertainty_eng.html). Moreover, in this paper, inputs of stochastic models are previously measured values, however, in a previous paper we have shown that exogenous data (parameters such as temperature, air pressure etc.) improve the prediction efficiency (Voyant et al., 2013b). As it was our first experiment in using geostationary satellites data, we have preferred to start without the multivariate case.

In the next section, the Meteosat images acquisitions, the SSI computation by Heliosat-II model described in (Rigollier et al., 2004; Gueymard, 2012) and clear sky index computing methodologies (Maini and Agrawal, 2006) will be first explained. Then we will detail the methodologies of prediction we have tested, taking care to explain first the TS formalism dedicated to the 2-D global solar radiation modeling. Then we will expose results of evaluation between modeling and HC-3 SSI over 2 years allowing the cross-comparison of models. Finally we will close the paper with a conclusion.

## 2. Materials and methods

Before to expose the time series modeling in the 2-D case, two brief sections are dedicated to the global radiation estimations from satellite acquisitions, the determination of clouds and the clear sky index described in (Şenkal, 2010; Singh et al., 2011; Thies and Bendix, 2011). The clear sky index based on the European Solar Radiation Atlas (ESRA) model (Rigollier et al., 2000) has the characteristic to define a TS made stationary (Cybenko, 1989; Hornik et al., 1989) and therefore directly usable by most stochastic models (as MLP for example).

### 2.1. Satellite-derived hourly radiation

Many methods allow a precise knowledge of spatial distribution and temporal behavior of SSI over a territory. Among these methods, remote sensing allows processing efficiency satellite images to estimate the cloud cover related to the studied area and so, indirectly the SSI. The interest of SSI estimated from satellite data for a great area has been demonstrated in (Muselli et al., 1998a; Perez et al., 2013). In this study, SSI was extracted from the Helioclim-3 database (Vernay et al., 2013) centered on Corsica. This database provides hourly SSI maps with a nadir spatial resolution of 2.5 km obtained from Meteosat Second Generation (MSG) images treated by the Heliosat-2 model (Guemene Dountio et al., 2010). This model uses calibrated satellite images that are converted into irradiance. They result from radiation-atmosphere interactions: a fluctuation of the signal measured by the sensors is interpreted as a variation of the cloud cover. Heliosat-2 computes the SSI from the clear sky radiation multiplied by a cloud index $n_t(x_i, y_j)$ which expresses the clearness of the atmosphere for the pixel $(x_i, y_j)$ and the time $t$ (equation 1).

$$n_t(x_i, y_j) = \frac{\rho_t(x_i, y_j) - \rho_t^{CS}(x_i, y_j)}{\rho_t^{cloud}(x_i, y_j) - \rho_t^{CS}(x_i, y_j)} \tag{1}$$

In this equation, $\rho_t(x_i, y_j)$ is the reflectance (or apparent albedo) measured by the sensor, $\rho_t^{cloud}(x_i, y_j)$ is the apparent albedo of the brightest clouds and $\rho_t^{CS}(x_i, y_j)$ is the apparent albedo of the ground under clear sky. (Rigollier et al., 2004) have explicated the relation between the $n_t(x_i, y_j)$ parameter and the clear sky index noted $CSI_t(x_i, y_j)$ as described in the equation 2.

$$\begin{aligned}
CSI_t(x_i, y_j) &= 1.2 \text{ if } n_t(x_i, y_j) < -0.2 \\
CSI_t(x_i, y_j) &= 1 - n_t(x_i, y_j) \text{ if } -0.2 < n_t(x_i, y_j) < 0.8 \\
CSI_t(x_i, y_j) &= 2.0667 - 3.6667 n_t(x_i, y_j) + 1.6667 \left(n_t(x_i, y_j)\right)^2 \text{ if } 0.8 < n_t(x_i, y_j) < 1.1 \\
CSI_t(x_i, y_j) &= 0.05 \text{ if } n_t(x_i, y_j) > 1.1
\end{aligned} \tag{2}$$

As the global radiation $I_t(x_i, y_j)$ and the global radiation under clear sky $I_t^{CS}(x_i, y_j)$ are linked by the simple relation $CSI_t(x_i, y_j) = I_t(x_i, y_j) / I_t^{CS}(x_i, y_j)$, it is particularly evident to generate $I_t(x_i, y_j)$ if $I_t^{CS}(x_i, y_j)$ is known. The clear sky model implemented in Heliosat-2 was developed in the framework of the European Solar Radiation Atlas (ESRA model) (Rigollier et al., 2000). In this model, the global horizontal irradiance under clear sky is split into direct and diffuse components. Their computations are based on the Linke turbidity factor (Remund et al., 2003) that summarizes the turbidity of the atmosphere, and hence the attenuation of the direct beam and the generation of the diffuse fraction (Kasten, 1996).

### 2.2. Studied zone and data

Corsica is a Mediterranean island located in the golf of Genoa. It stretches out on 183 km long between latitudes 41° 20' 02" N and 43° 01' 31" N and 83.5 km wide between longitudes 8° 32' 30" E and 9° 33' 38" E extending to an area of 8722 km$^2$. The island's seaside globally benefits from a Mediterranean climate characterized by hot dry summers and mild wet winters. Corsica presents one of the greatest solar potential of metropolitan France. The island's mountainous inner has a climate of height elevations or even alpine climate. The summers are hot and sunny but the weather is more instable on winters and the precipitations are abundant. Thus, the geographical distribution of solar potential and the dynamics of meteorology are particularly heterogeneous in such a territory. The public meteorological network is composed of only six stations. Only three of them provide hourly radiations measurements while the others provide 10 days integrated data (Haurant et al., 2012). In this context, predictions from 1-D TS modeling can be realized for three punctual locations in Corsica, not enough to allow a global management of electrical network and photovoltaic plants. Thus 2D prediction map from satellite SSI maps can be a good alternative related to the lack of measured data. In the present study, 1158 points are considered, defined by the HelioClim-3 meshgrid centered on Corsica (not necessarily coincident with the native MSG pixels). Figure 1 shows this meshgrid and three of the principal cities of the island where there are weather stations. In fact, each SSI pixel is a generator of one global radiation time series. Eight years of HC-3 hourly global radiations are available, from 2005 to 2012 for these three places. The two last years (2011-2012) are used for testing the different predictors and the six first years (2005-2010) are dedicated to the learning phase concerning the MLP approach.

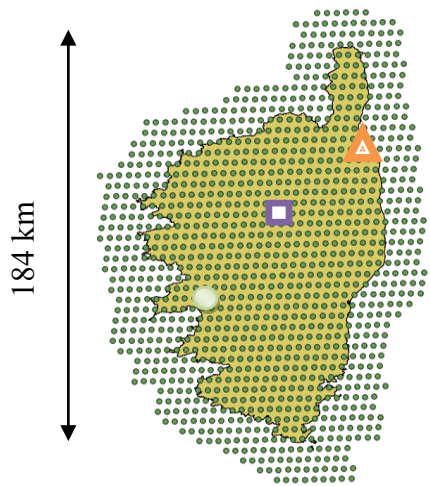

Figure 1. Points of the HelioClim3 meshgrid and location of the 3 meteorological stations providing hourly solar radiation measurements, circle locates Ajaccio (41°55'N and 8°44'E, elev. 0-787 m), square locates Corte (42°18'N and 9°09'E, elev. 300-2626 m) and triangle locates Bastia (42°42'N and 9°27'E, elev. 0-963 m)

## 2.3. Prediction models

There are a lot of different approaches to model time series and comprehensive reviews are available (Hamilton, 1994; Mellit et al., 2009; Şahin et al., 2012.; Voyant et al., 2013). The common base of all these models is that an element of TS ($I_{t+1}(x_i, y_j)$) can be defined by a linear or non-linear model called $f_n$ (see Equation 3 where $t = n, n-1, ..., p+1, p$ with $n$, the number of observations and $p$ the number of parameters with $n \gg p$, and $\epsilon_{t+1}(x_i, y_j)$ the error term of the prediction à time $t+1$ and for the pixel $(x_i, y_j)$) (Crone, 2005).

$$I_{t+1}(x_i, y_j) = f_n(I_t(x_i, y_j), I_{t-1}(x_i, y_j), \ldots, I_{t-p+1}(x_i, y_j)) + \epsilon_{t+1}(x_i, y_j) \tag{3}$$

In the extended territory case, the previous equation is applicable for each pixel of the image as described in the Figure 2. The following part is dedicated to particular families of functions $f_n$, especially related to the persistence and its scaled form, artificial neural networks and to the cloudless approximation and modeling. Note that the ARMA approach is not tested here, because in a previous study (Voyant et al., 2013b) we show that MLP and ARMA give the same results for the global radiation prediction in Corsica Island in the univariate case.

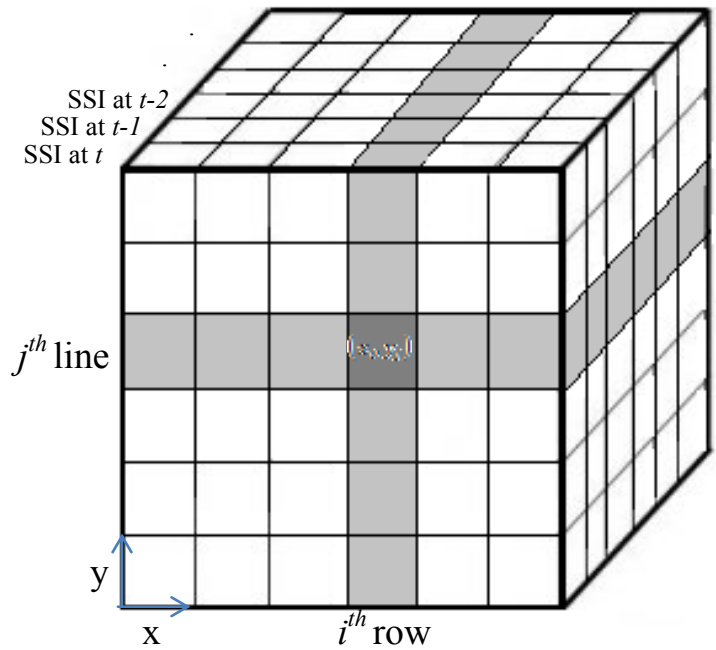

Figure 2. TS definition from successive satellite acquisitions. In dark grey the pixel $(x_i, y_j)$ at time $t$ defining the global radiation noted $I_t(x_i, y_j)$. The intensity of successive pixels defines each TS ( $\ldots I_{t-2}(x_i, y_j), I_{t-1}(x_i, y_j) I_t(x_i, y_j) \ldots$).

### 2.3.1. Persistence and scaled persistence

The first of studied prediction methods is the persistence method; the simplest way of producing a forecast. The persistence method assumes that the atmospheric conditions are stationary as shown in the equation 4 (Voyant et al., 2013a).

$$\hat{I}_{t+1}(x_i, y_j) \xrightarrow{pers} I_t(x_i, y_j) \tag{4}$$

To take into account the fact that the apparent position of the sun is not identical between *t* and *t+1*, it is possible to correct the previous form with a clear sky ratio term (see equation 5) (Sfetsos and Coonick, 2000) and to generate the scaled persistence.

$$\hat{I}_{t+1}(x_i, y_j) \xrightarrow{scaled\ pers} I_t(x_i, y_j) \cdot \frac{I_{t+1}^{CS}(x_i, y_j)}{I_t^{CS}(x_i, y_j)} \tag{5}$$

### 2.3.2. Clear sky approximation

Another way to estimate the global radiation the next hour is to consider that we are dealing exclusively with sunny days (without cloud cover). In this case, the prediction is done only with determinist model as clear sky model presented previously (see equation 6) (Rigollier et al., 2000).

$$\hat{I}_{t+1}(x_i, y_j) \xrightarrow{CS} I_{t+1}^{CS}(x_i, y_j) \tag{6}$$

### 2.3.3. Artificial neural networks

To estimate the global radiation with stochastic models, a stationary hypothesis is often necessary. This result, initially shown for autoregressive and moving average processes (Hamilton, 1994) can be also applicable for the study and prediction from artificial neural networks (ANN) (Mellit et al., 2009). In fact, before to use ANN, time series must be made stationary. In previous work we have demonstrated that the clear sky index is a stationary computed series that can be directly operated by neural networks (Voyant et al., 2012).

Although a large range of different architecture of ANN is available, MultiLayer Perceptron (MLP) remains the most popular (Mellit et al., 2009). In particular, feed-forward MLP networks with two layers (one hidden layer and one output layer) are often used for modeling and forecasting time series. Several studies have validated this approach based on ANN for the non-linear modeling of time series. To forecast the next value of a time series, a fixed number *p* of past values are set as inputs of the MLP, the output is the prediction of the future value (Voyant et al., 2012). Considering the initial time series equation (equation 3), we can transform this formula in the non-linear case of one hidden layer MLP with *b* related to the biases, *f* and *g* to the activation functions of the output and hidden layer, and $\omega$ to the weights (see equation 7). The number of hidden nodes (*H*) and input nodes (*In*) allow detailing this transformation. Note that as MLP is a stationary estimator, it is the clear sky index which is used during the modeling which defines a time series made stationary. To build a predictor of the global radiation, it is possible to construct and train a MLP with the clear sky estimation at time t+1 and the clear sky index of previous lags.

$$\hat{I}_{t+1}(x_i, y_j) = I^{CS}_{t+1}(x_i, y_j) f\left(\sum_{i=1}^{H} o_i \, \omega_i^2 + b^2\right)$$

$$\text{With } o_i = g\left(\sum_{j=1}^{In} CSI_{t-j+1}(x_i, y_j)\, \omega_{ij}^1 + b_i^1\right) \tag{7}$$

In the presented study, the MLP has been computed with the Matlab© software and its Neural Network toolbox. The characteristics chosen and related to previous works (Voyant et al., 2013b): one hidden layer, the activation functions are the continuously and differentiable hyperbolic tangent (hidden) and the linear function (output), the Levenberg-Marquardt (approximation to the Newton's method) learning algorithm with a max fail parameter before stopping training equal to 3 (Voyant et al., 2013b). The detail of the MLP architecture used is shown in the Figure 3. In fact a similar MLP is defined for all the pixels of the radiation map: 1158 different MLP without any hypothesis on the spatial coherency.

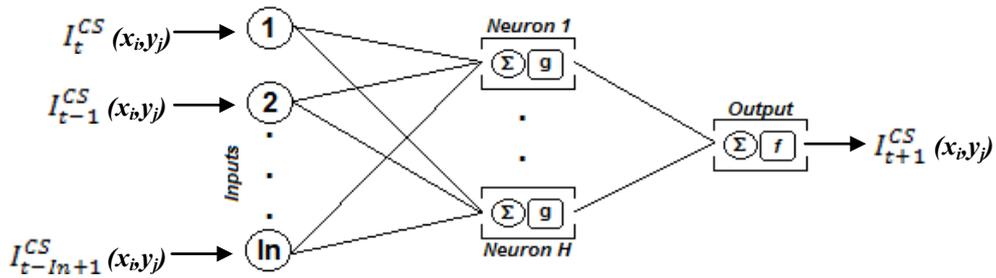

Figure 3. MLP architecture for the pixel $(x_i, y_i)$

Training, validation and testing data sets were respectively set to 80%, 20% and 0% (Matlab parameters). These three phases concern the six first years (4.6 years for training and 1.2 years in order to execute an early stopping), and the two last years (2011-2012; not use during the training) are used for the global solar radiation forecasting test. All predictors are compared during these two last years. In order to optimize the different MLP, it is essential to construct a tool able to determine the number and the nature of the MLP inputs. We propose to use the automutual information tool ($MI(X_t, L^\tau X_t)$ for $\tau > 1$ in equation 8) which is a quantity measuring the automutual dependence of a variables $X_t$ (*L* and $\tau$ the lag operator and associated order). In fact, this formalism replaces and generalizes the auto-correlation which allows to measures only the linear relationship. Mutual information is more general and measures the reduction of uncertainty in $L^\tau X_t$ after observing $X_t$ (Parviz et al., 2008). So *MI* (Jiang et al., 2010) can measure non-monotonic and other more complicated relationships. It can be expressed as a combination of marginal and conditional entropies (respectively $H(X_t)$ and $H(X_t|L^\tau X_t)$) as described in the equation 8 (Huang and Chow, 2005).

$$MI(X_t, L^\tau X_t) = H(X_t) - H(X_t|L^\tau X_t) \qquad (8)$$

Entropy corresponds to a measure of unpredictability or information content and can be written by the expression detailed in the equation 9 (entropy of a discrete random variable $X_t$ with possible values $x=\{x_1,...x_n\}$).

$$H(X_t) = -\sum_x p(x) \log(p(x)) \qquad (9)$$

One may also define the conditional entropy of two events $X_t$ and $L^\tau X_t$ (Equation 10). This quantity should be understood as the amount of randomness of the random variable $X_t$ given that you know the value of $Y_t = L^\tau X_t$ (possible values $y=\{y_1,..y_n\}$).

$$H(X_t|L^\tau X_t) = -\sum_x \sum_y p(x,y) \log(p(x)/p(x,y)) \qquad (10)$$

The definition of the joint probability distribution function ($p(x,y)$) and marginal probabilities ($p(x)$ and $p(y)$), allows to define a new form of the mutual information as described in the equation 11.

$$MI(X_t, L^\tau X_t) = \sum_x \sum_y p(x,y) \log\left(\frac{p(x,y)}{p(x)p(y)}\right) \qquad (11)$$

We consider that the maximum lag to take as input of the MLP corresponds to the first minimum of the automutual information graph ($In = \mathrm{argmin}_\tau MI(CSI_t, L^\tau CSI_t)$). In example presented in the Figure 4 (the *MI* versus the time lag for a pixel near Ajaccio), there is a first minimum corresponding to the 10th time lag, and thereby inducing a MLP with 10 inputs ($I_t^{CS} ... I_{t-9}^{CS}$).

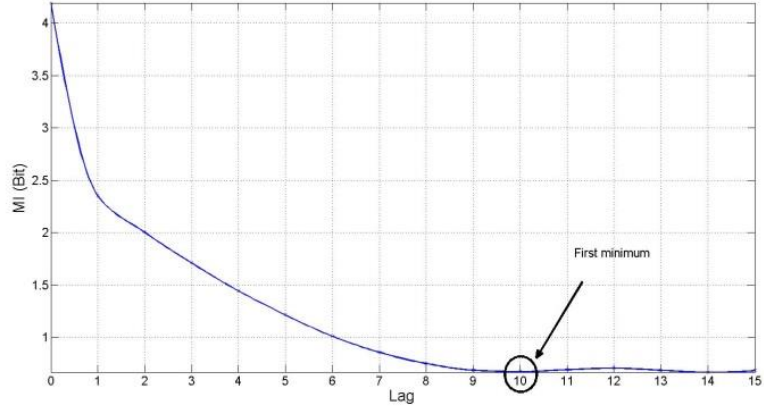

Figure 4. Example of automutual information of the clear sky index in the Ajaccio site

The first minimum of auto-mutual information concerning all the points of the meshing (1158 different operations) has a median value equal to 7 time lags (min=5, max=10, mean value=7.63 and standard deviation=1.08). Thereby, for the overall territory, the number of inputs (In) is takes equal to 7 (i.e. 7 hours of passed measurements). Related to previous studies (Voyant et al., 2013a, 2012), we consider that the optimal configuration of the MLP corresponds to one hidden layer provided with 7 hidden neurons (size of inputs and hidden nodes should be equivalent).

## 3. Results

Before exposing prediction results, it is important to define error metrics. During this study, two parameters are dedicated to the standard TS analysis (see Equations 12 and 13) and one has been constructed to take into account the 2-D aspect of predictions and measurements (Equation 14 and 15). The nRMSE metrics is expressed with the expected value ($E$) which is in this case, the temporal mean computed on the overall test sampling.

$$\epsilon_{t+1}(x_i, y_j) = I_{t+1}(x_i, y_j) - \hat{I}_{t+1}(x_i, y_j) \qquad (12)$$

$$nRMSE = \sqrt{E\left[\left(I_{t+1}(x_i, y_j) - \hat{I}_{t+1}(x_i, y_j)\right)^2\right]} / E[I_t(x_i, y_j)] \qquad (13)$$

In the present study, the clear sky index is used in all the predictive methodologies and is related to some geographical (altitude, longitude or latitude) and modeling parameters (Linke turbidity factor or relative air mass). Because of the sampling process of the HC-3 SSI, it is possible that for some pixels, these parameters could not be adapted (especially altitude, if it is in a mountainous region). In this context, we need a tool giving information about the fact that neighbor pixels could have a better forecasted value. In physics and particularly in medical physics, the gamma index is a criterion for comparing data from two matrixes. This interpretation is based on the assumption that sometime the discrepancy between measures and calculi from the model may not be due to an error of the map calculation algorithm or delivery hardware, but simply due to experimental error (different offsets or spatial resolutions). In the high gradient region, a small offset could induce a significant error between real pixel and forecasted one while the modeling is good. More abstractly, the gamma index takes two 2D arrays and compares each element of array_1 with spatially-nearby elements of array_2. The gamma test was first introduced by Low et al. (1998) as a single metric that combined features of both pixel intensity difference (PID; as describe in equations 12 or 13) and distance to agreement (DTA; e.g. distance between iso-intensity curves), allowing a robustly performing in the regions prone to failure. It is the minimum Euclidean distance in a normalized intensity pixel-distance space. In this way, both PID and DTA are taken into account for every point in the evaluated distribution (Thomas and Cowley, 2012) generating a dimensionless metric. For a measured image (*m*) and predicted image (*p*), the gamma index ($\gamma$) is defined from the distance (noted $r(r_p, r_m)$, Equation 14) between points in reference (or measurement; $r_m$) and evaluation distributions (or prediction ; $r_p$), note that polar coordinates are used but a Cartesian approach is also possible:

$$r(r_p, r_m) = |r_p - r_m| \tag{14}$$

If $\Delta I(r_p, r_m)$ is the pixel intensity difference at point $r_p$ in the predicted image and point $r_m$ in the measured image ($\Delta I(r_p, r_m) = I_p(r_p) - I_m(r_m)$), so the gamma index ($\gamma$) is defined for two tolerance terms ($Tol_r$ for the distance and $Tol_I$ for the pixel intensity) by the equation 15 ($Tol_r$ is in fact a fixed DTA and $Tol_I$, a fixed PID). The minimum radial distance between the measurement point and the calculation points (expressed as a surface in the intensity–distance space of the Figure 5) induced the $\gamma$ index metrics.

$$\gamma = \min\{\Gamma(r_p, r_m)\} \text{ for all } r_p \text{ where } \Gamma(r_p, r_m) = \sqrt{\frac{r^2(r_p, r_m)}{Tol_r^2} + \frac{\Delta I^2(r_p, r_m)}{Tol_I^2}} \tag{15}$$

Each pixel of the image is associated with a gamma index value. The higher is this index, the less the prediction is good so the determination of $\gamma$ throughout the predicted irradiance distribution provides a presentation that quantitatively indicates the calculation accuracy. Moreover, regions where $\Gamma > 1$ correspond to locations where the calculation does not meet the acceptance criteria. Note that, it possible to construct the same type of test with the hypothesis that $Tol_r \to +\infty$, in this case the physical distance is not taken into account, only the pixels intensity are considered ($\Gamma(r_p, r_m) = \frac{\Delta I\ (r_p, r_m)}{Tol_I}$).

With this index, it is possible to construct a statistical test based on the null hypothesis H0: the prediction related to the pixel $r_p$ is relevant. If $\gamma(r_m) > 1$, H0 is rejected else H0 is accepted. This test is represented in the Figure 5 for each pixel in a space composed of pixel intensity and spatial coordinates. The acceptance criteria form an ellipsoid surface, the major axis scales of which are determined by individual acceptance criteria and the center of which is located at the estimated point in question. When the calculated intensity distribution surface passes through the ellipsoid, the calculation passes the acceptance test for the measurement point.

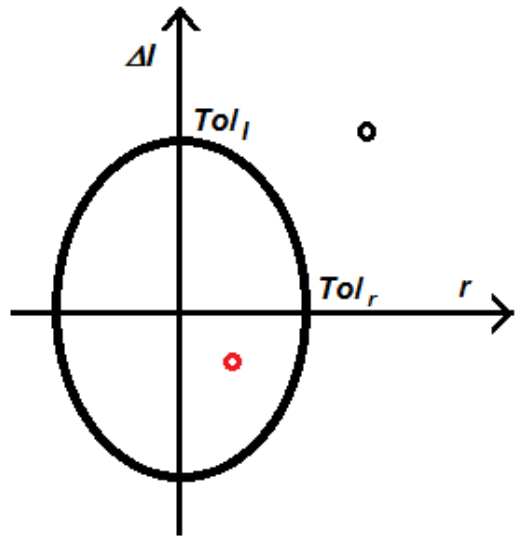

Figure 5. Schematic representation of the gamma index test. Inside the ellipse, the test is accepted (red point) and outside, it is rejected (black point).

The measure of acceptability is the multidimensional distance between the measurement and calculation points in both the pixel intensity and the physical distance, scaled as a fraction of the acceptance criteria. For an overall image, it is possible to synthetize the test computing the gamma passing rate (%GP; for each map it is the percentage of pixels passing the test), higher is this percentage and better is the prediction.

## 3.1. Performances of HC-3 SSI estimations

The data used during this study are related to global solar estimation from the Helioclim-3 database. The performances of the satellite derivate SSI estimations have been evaluated in the case of Corsica. In that view, they have been compared to hourly radiometric measurements provided by three meteorological stations in Corsica (Fig 1, table 1) that cover the period from April 1$^{st}$, 2004 to December 17$^{th}$, 2006.

The table 1 summarizes seasonal performances of SSI estimations for the three stations. It is shown these performances when irradiations under 0.1 Wh/m² and 10 Wh/m² are eliminated. In the first case all the night values are removed (~6:00-22:00 in summer; period corresponding to daytime) and concerning the second case, one or two additional hours are removed. Imposing such a threshold allows limiting uncertainties due to the air optical air mass artifacts at sunrise and sunset and shading effects.

| Periods | Threshold | Ajaccio | Corte | Bastia | Mean |
|---|---|---|---|---|---|
| Year | 0.1 | 19.8 | 23.5 | 22.0 | 21.8 |
| | 10 | 18.3 | 22.6 | 19.5 | 20.1 |
| Winter | 0.1 | 24.5 | 29.6 | 27.6 | 27.2 |
| | 10 | 23.6 | 28.0 | 23.0 | 24.8 |
| Spring | 0.1 | 16.4 | 20.8 | 20.9 | 19.4 |
| | 10 | 15.7 | 20.2 | 18.1 | 18.0 |
| Summer | 0.1 | 18.7 | 20.0 | 19.5 | 20.1 |
| | 10 | 17.9 | 19.5 | 17.7 | 18.3 |
| Autumn | 0.1 | 19.3 | 31.9 | 22.9 | 24.7 |
| | 10 | 18.6 | 29.9 | 21.6 | 23.4 |

Table 1. Comparison between helioclim-3 satellite SSI and ground measurements for the 3 studied stations in Corsica (nRMSE in %)

Firstly, we can see that the 10 Wh/m² threshold allows decreasing the errors of 0.5 to 1.5 points demonstrating that the sunrise and sunset artefacts induce important errors, especially in Corte (the only mountainous site) but also in Bastia. Don't forget that the nRMSE is also linked with errors in the reference data: the location of radiometric stations is not necessarily relative to the center of each pixel. We can also observe that the cloudy periods (winter - autumn) present the most important error (nRMSE between 20 and 30 %) showing that SSI under cloudy sky are particularly difficult to be calculated by the model as we could have expected. However, during cloudy months the increasing of observed nRMSE can be also or partly explained by a decreasing of the reference irradiation (see equation 13). In fact, the HelioClim3 database (due to the pixel size) is not very sensitive to small clouds formations while they can have a big impact on the ground measures. Finally we can note that the estimations are correlated with measurements since yearly and seasonal cross correlation coefficients are over 0.96. Yearly nRMSE are between 18.0 and 27.2 %, comparable but lightly over the nRMSE of Meteosat-8 estimations for the same stations and the same periods (Haurant et al., 2012). However, these data have been selected for this study because HelioClim-3 is one of the databases that offer SSI maps (or data series) on a long period but others models exist (CM-SAF, Land-SAF, etc.) as described in (Ineichen et al., 2009). Also, the HelioClim-3 extracted maps have higher resolution than Meteosat-8 images. HelioClim-3 allows also the possibility of near real-time SSI acquisitions: the day after in standard mode and on specific requests, 15 minutes MSG acquisitions plus some minutes on a selected region.

## 3.2. Irradiation prediction validation

Now the sensibility study is done concerning the validation of the HC-3 data, in this part we will expose the performance of the four methods of prediction described in the section 2. During all the experimentations only hours between 8:00 and 18:00 are used (UTC), outside this interval the choice to predict the global radiation is not really justified. The table 2 describes the comparison between the four predictors. The nRMSE, the gamma index and the gamma passing rate are given for the four seasons. The gamma index is related to a $Tol_r$ = 1 pixel (2.5 km) and $Tol_I$ = 10%. We choose intuitively these values because this parameter is never used in global radiation forecasting. If gamma is less than 1 it corresponds to a prediction error lower than 10% with a precision better than 2.5 km.

*gamma passing rate

| Criteria | Scaled persistence | MLP | Persistence | Clear sky model |
|---|---|---|---|---|
| nRMSE (%) | 16.61 | **16.54** | 34.11 | 45.04 |
| $\gamma_{wi}$ | **0.84** | 1.21 | 1.21 | 4.9 |
| $\gamma_{sp}$ | **0.36** | 0.94 | 1.3 | 1.01 |
| $\gamma_{su}$ | **0.04** | 0.27 | 0.28 | 0.12 |
| $\gamma_{au}$ | **0.27** | 0.58 | 0.51 | 0.23 |
| %GP$_{wi}$* | **67.7** | 45.6 | 47.6 | 0.01 |
| %GP$_{sp}$ | **86.0** | 58.4 | 12.4 | 78.4 |
| %GP$_{su}$ | 99.5 | **99.8** | 99.7 | 98.6 |
| %GP$_{au}$ | 93.8 | 94.5 | **95.3** | 94.5 |

Table 2. Comparison of the 4 predictors, the best values are in bold

The two better predictors are the scaled persistence and the MLP, their nRMSE is less than 17%, the average gamma index is lower than 1 (except 1.21 for MLP in spring) and the gamma passing rate (%GP) is upper than 65% for the scaled persistence and upper than 45% for MLP.

Concerning the clear sky model, the gamma passing rate is very good in summer (more than 98%; related to few cloud occurrence) and very bad in winter (less than 1%; related to a lot of cloud occurrences). Note that the variability of the global radiation from satellite acquisitions is less important than in the case of ground-based measurements. If in the first case the nRMSE of the prediction is close to 20% and in the second it is less than 17%. In fact, the "small clouds" are not well considered in the Helioclim-3 data (because of the pixels size), there is sometime overestimation and then underestimation regarding ground measurements. As scaled persistence and MLP predictions are equivalent, the simplicity of the first one, led us to think it is the most interesting. In the Figure 6 the gamma test is presented for 4 typical days related to each season with the scaled persistence. Black pixels if the test is rejected and white if it is passed.

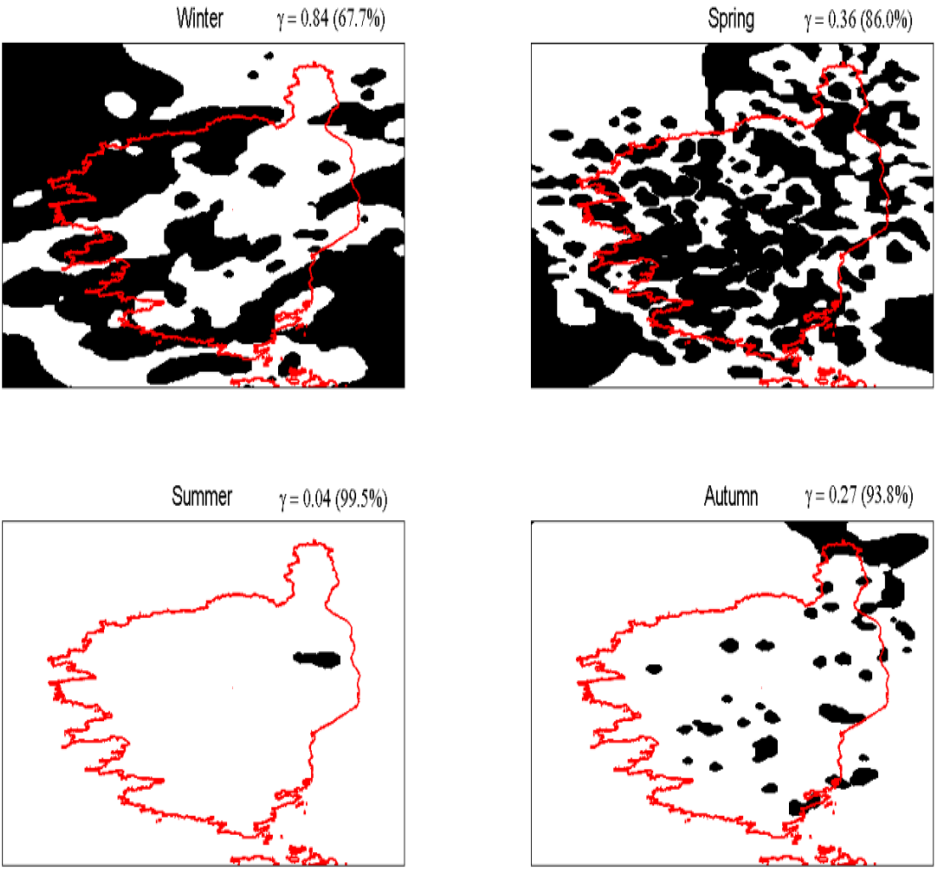

Figure 6. Gamma test for 4 typical days with scaled persistence. Black pixel if the test is rejected and white pixel if it is passed

During summer and autumn, there are a lot of white pixels, almost all the predictions are in the ellipse 2.5km/10%. For the other seasons, results are less uniform: in spring the number of black and white pixels is equivalent but in winter, there are more black than white pixels. Another way to see the quality of the prediction is to generate the algebraic error ($\epsilon_{t+1}(x_i, y_j)$) between prediction (for typical days) and measurement as shown in figure 7. If in winter the error is significant (a lot of red pixels indicating that absolute error is upper than 100Wh/m²), in summer it is close to 0Wh/m² (a lots of light blue pixels). In this example, there is not significant relationship between the altitude and the error of prediction (mountain ridge in the center of the island longitudinally).

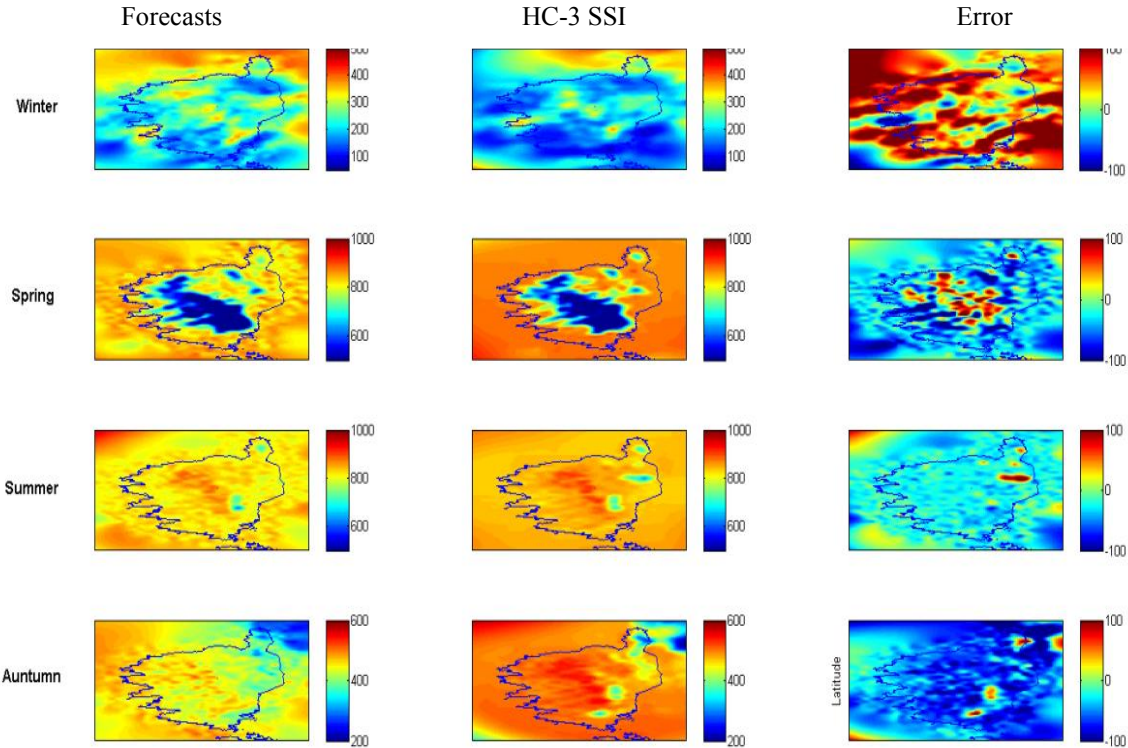

Figure 7. Forecasted SSI with scaled persistence, HC-3 SSI and algebraic error $\epsilon_{t+1}(x_i, y_j)$ (Color map in Wh/m²)

In the Figure 8a and b are represented the spatial repartition of nRMSE and the histogram of the global nRMSE of the best predictor (scaled persistence). The center of the Gaussian fit is 16.6%. The behavior of the nRMSE seems to follow a normal distribution reminiscent of a simple random walk (no bias and no strong heterogeneity in the nRMSE). Note that the high altitude pixels give better result than peripheral regions (in blue in the middle of Corsica), and that there is an asymmetry between east and west coasts.

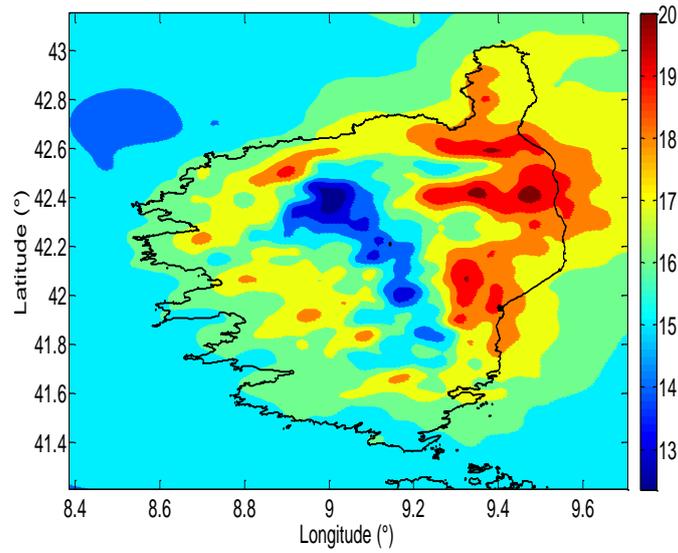

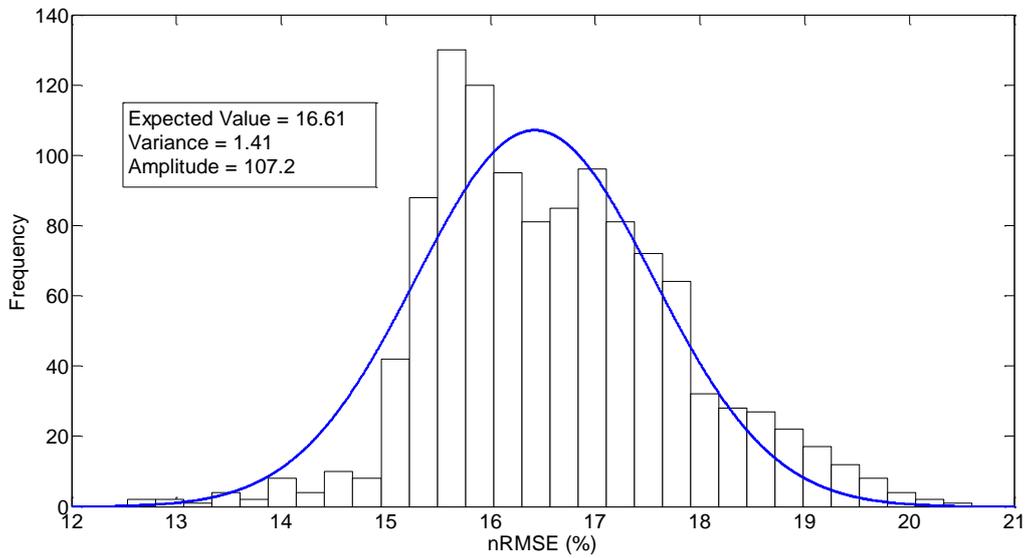

Figure 8. Distribution of the annual *nRMSE* (*a.* spatial repartition and *b.* histogram)

To appreciate visually the quality of prediction, it is essential to draw profiles of HC-3 SSI map and predictions as in the Figures 9-11. In these plots we see that during the summer period the prediction is effective unlike the other seasons.

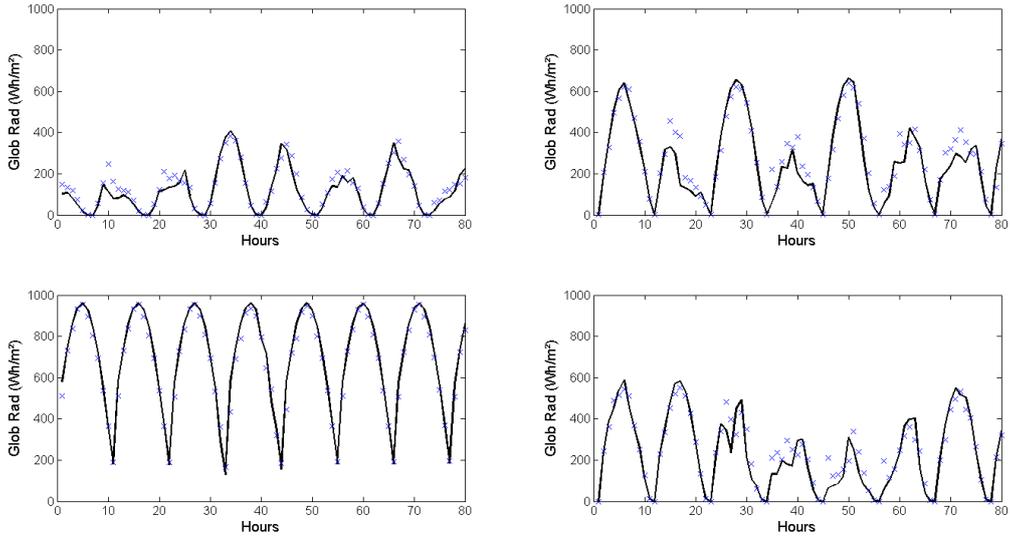

Figure 9. Profiles of HC-3 SSI data and scaled persistence estimation of the global radiation (respectively line and marks) in a pixel localized near Ajaccio

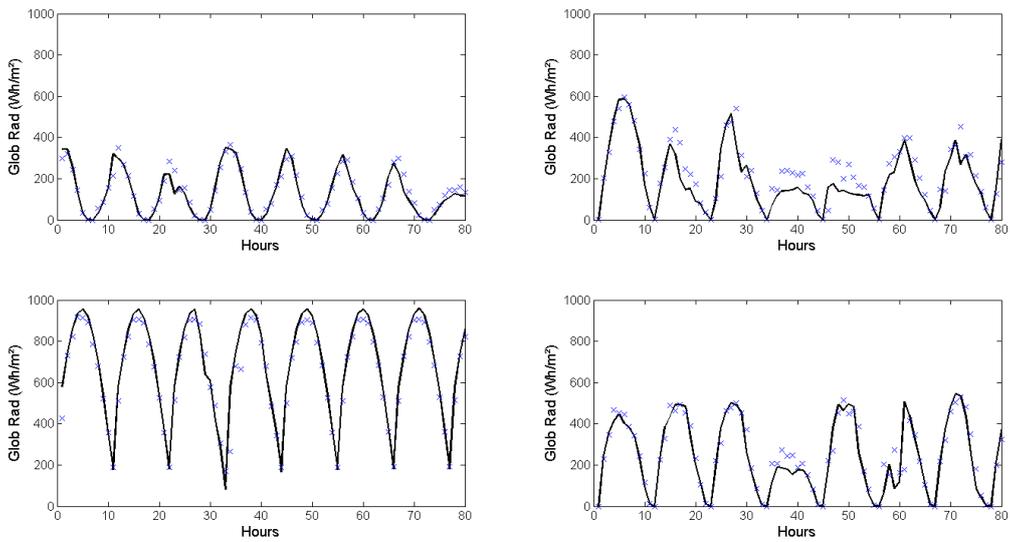

Figure 10. Profiles of HC-3 SSI data and scaled persistence estimation of the global radiation (respectively line and marks) in pixel localized near Corte

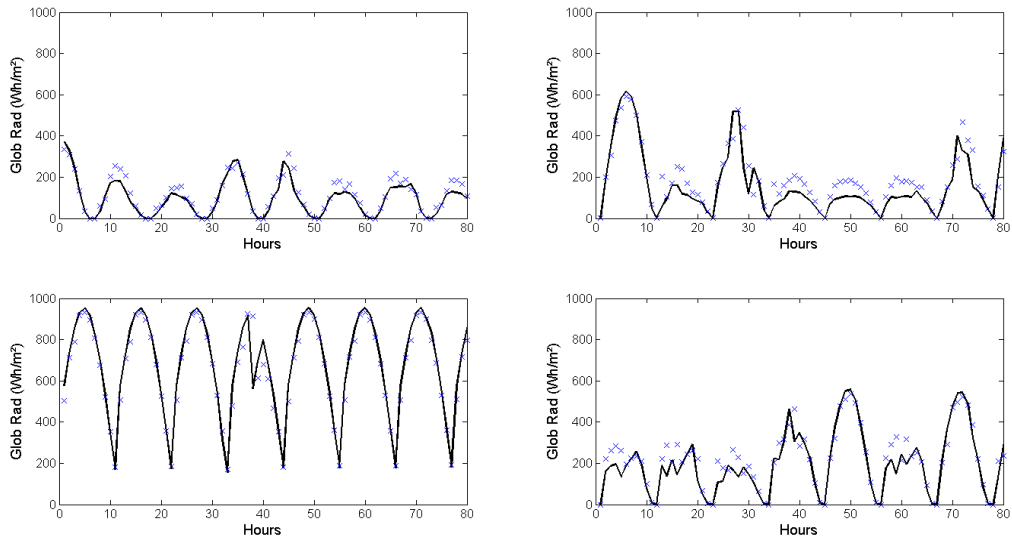

Figure 11. Profiles of HC-3 SSI data and scaled persistence estimation of the global radiation (respectively line and maks) in pixel localized near Bastia

## 4. Conclusions

With this article, we add a new original approach for global radiation prediction. A methodology usually applied for 1-D time series modeling has been modified and used to generate a 2-D predictions map for an overall territory (Corsica Island in our case). Classical time series analyze tools were used, however a lot of new methods are developed:

-usually, the TS formalism is applied for a specific location and not covering an overall spatial territory (resolution of 2.5km);

-the mutual information as MLP optimization parameter is rarely used;

-the scaled persistence is very rarely used in time series prediction;

-the gamma index is never used in global radiation forecasting.

The methodology applied admits two sources of uncertainty, the first one is linked to the Helioclim-3 database and the global radiation computing from satellite acquisitions, the second one is linked to the prediction method used. The value of nRMSE including the two error components is close to 37% (respectively 20.1% and 16.5%). Concerning the gamma passing rate, our study shows that for winter and spring the scaled persistence gives the best results (respectively 67.7% and 86%), for autumn, it is the simple persistence (95.3%) and for summer it is the MLP (99.8%). We think that this new methodology based on the 2-D time series modeling with scaled persistence (or MLP) has to be compared against the complex model NWP (with different pixels sizes and origins offsets). The next step will be to compare in Corsica, using the gamma index and nRMSE, the prediction of NWP reference methodologies (NOAA and AROME) and the time series approaches (MLP and scaled persistence). Moreover, if in the 1-D case the MLP gives very good result, in the 2-D case the interest of this methodology is not shown in this paper, new adaptations are necessary (specialized MLP by geographical clustering, use of adjacent pixels values, round off the optimization of each MLP, etc.). Finally, with MLP there is the operational possibility to integrate to the forecast schema, existing real-time measurements of irradiation and/or PV production at different locations in the region of interest. This approach which includes a possible calibration with historical existing databases and auto-correction in the real-time basis is certainly a good way to dramatically decrease the nRMSE of the forecast (Kalman filter, recursive least squares filter, etc.). Moreover, in this paper, inputs of models are previously measured values, the multivariate case (with exogenous data as inputs: wind speed, temperature, humidity, etc.) could improve the result and it will be a good perspective of future work.

.

# 5. Acknowledgement

Thanks to Armines and to the SoDa community for making the HelioClim data available (http://www.helioclim.org/index.html).

# List of captions

Table 1. Comparison between helioclim-3 satellite SSI and ground measurements for the 3 studied stations in Corsica (nRMSE in %)

Table 2. Comparison of the 4 predictors, the best values are in bold

Figure 1. Points of the HelioClim3 meshgrid and location of the 3 meteorological stations providing hourly solar radiation measurements, circle locates Ajaccio (41°55'N and 8°44'E, elev. 0-787 m), square locates Corte (42°18'N and 9°09'E, elev. 300-2626 m) and triangle locates Bastia (42°42'N and 9°27'E, elev. 0-963 m)

Figure 2. TS definition from successive satellite acquisitions. In dark grey the pixel $(x_i, y_j)$ at time $t$ defining the global radiation noted $I_t(x_i, y_j)$. The intensity of successive pixels defines each TS ($...I_{t-2}(x_i, y_j), I_{t-1}(x_i, y_j) I_t(x_i, y_j)...$).

Figure 3. MLP architecture for the pixel ($x_i, y_i$)

Figure 4. Example of automutual information of the clear sky index in the Ajaccio site

Figure 5. Schematic representation of the gamma index test. Inside the ellipse, the test is accepted (red point) and outside, it is rejected (black point).

Figure 6. Gamma test for 4 typical days with scaled persistence. Black pixel if the test is rejected and white pixel if it is passed

Figure 7. Forecasted SSI with scaled persistence, HC-3 SSI and algebraic error $\epsilon_{t+1}(x_i, y_j)$ (Color map in Wh/m²)

Figure 8. Distribution of the annual *nRMSE* (*a.* spatial repartition and *b.* histogram)

Figure 9. Profiles of HC-3 SSI data and scaled persistence estimation of the global radiation (respectively line and marks) in a pixel localized near Ajaccio

Figure 10. Profiles of HC-3 SSI data and scaled persistence estimation of the global radiation (respectively line and marks) in pixel localized near Corte

Figure 11. Profiles of HC-3 SSI data and scaled persistence estimation of the global radiation (respectively line and maks) in pixel localized near Bastia